\documentclass[3p,times]{elsarticle}

\usepackage{ecrc}

\volume{00}

\firstpage{1}

\journalname{Nuclear Physics A}

\runauth{P. Christiansen for the ALICE collaboration}

\jid{nupha}

\usepackage{amssymb}

\usepackage[figuresright]{rotating}
\usepackage{amsmath,amsfonts}
\usepackage{xspace}
\usepackage{upgreek}

\newcommand{\snnt}[1]{\ensuremath{\sqrt{s_{NN}} = #1 \text{ TeV}}\xspace}

\newcommand{\sppt}[1]{\ensuremath{\sqrt{s} = #1 \text{ TeV}}\xspace}

\newcommand{\gevc}[1]{\ensuremath{#1 \text{ GeV/$c$}}\xspace}
\newcommand{\dndeta}[1]{\ensuremath{\frac{\text{d}^2N_{#1}}{\text{d}\pt \text{d}\eta}}\xspace}

\newcommand{\dndypbpb}{\ensuremath{\left(\frac{d^2N}{d\pt dy}\right)_{\text{\pbpb}}}\xspace}

\newcommand{\dndsigmapp}{\ensuremath{\left(\frac{d^2\sigma_{\rm{INEL}}}{d\pt dy}\right)_{\text{pp}}}\xspace}
\newcommand{\eff}[1]{\ensuremath{\epsilon_{#1}}\xspace}

\newcommand{\yield}[1]{\ensuremath{Y_{#1}}\xspace}
\newcommand{\ch}{\ensuremath{\text{ch}}\xspace}
\newcommand{\sumpi}{\ensuremath{\uppi^- + \uppi^+}\xspace}
\newcommand{\ssumpi}{\ensuremath{\uppi}\xspace}
\newcommand{\sumkp}{\ensuremath{\text{K}^- + \text{K}^+ + \bar{\text{p}} + \text{p}}\xspace}
\newcommand{\ssumkp}{\ensuremath{\text{K} + \text{p}}\xspace}

\newcommand{\VZ}{\ensuremath{V^{0}}\xspace}

\newcommand{\RAA}{\ensuremath{R_{\text{AA}}}\xspace}
\newcommand{\TAA}{\ensuremath{T_{\text{AA}}}\xspace}

\newcommand{\pp}{pp\xspace}
\newcommand{\pbpb}{Pb-Pb\xspace}

\newcommand{\pt}{\ensuremath{p_{\text{T}}}\xspace}
\newcommand{\dpi}{\ensuremath{\Delta_{\pi}}\xspace}

\newcommand{\dedx}{\ensuremath{\text{d}E/\text{d}x}\xspace}
\newcommand{\mdedx}{\ensuremath{\left <\text{d}E/\text{d}x \right>}\xspace}

\newcommand{\mdedxpi}{\ensuremath{\left <\text{d}E/\text{d}x \right>_{\pi}}\xspace}

\newcommand{\LA}{\ensuremath{\Uplambda}\xspace}

\newcommand{\KOs}{\ensuremath{\text{K}^0_s}\xspace}

\begin{document}

\begin{frontmatter}



\dochead{}

\title{High \pt identified particle production in ALICE.}


\author{P. Christiansen for the ALICE Collaboration}

\address{Division of Particle Physics, Lund University, Sweden}

\begin{abstract}
The ALICE experiment is a dedicated heavy ion physics detector at the LHC with
unique capabilities for studying identified particle production. In this
proceeding preliminary results for \RAA for $\uppi$ and K+p (sum), are reported, based on measurements in \pp at
\sppt{2.76} and Pb-Pb at \snnt{2.76}. The results are compared to theoretical
predictions and measurements at RHIC.
\end{abstract}

\begin{keyword}

LHC \sep ALICE experiment \sep spectra \sep identified particle production \sep high pt \sep RAA

\end{keyword}

\end{frontmatter}


\section{Introduction}

The production of particles at high \pt in \pp collisions can be described
using perturbative QCD. In Pb-Pb collisions these hard probes are important
tools for studying the medium formed, as the initial state production can be
established from pQCD and binary scaling of \pp results. 

The observed yield of high \pt particles is much smaller than expected from
binary scaling because of strong final state interactions with the surrounding
dense medium~\cite{Aamodt:2010jd}. Experiments at RHIC have shown that this
modification is very different for mesons and
baryons~\cite{Adcox:2003nr,Adams:2006wk,Abelev:2006jr}.

The results from RHIC has lead to theoretical speculations on particle specie
dependent effects (PID effects for short) at high \pt that are extremely
attractive to test at LHC where the production cross section for hard
processes is much larger than at RHIC energies. In the following we shall
discuss 3 regimes of \pt (low, intermediate, high) and their PID effects in
\pbpb collisions.

The main PID effect at low \pt, $\pt < \gevc{2}$, is flow. For hydrodynamic
flow the PID dependence is purely due to mass differences (but the final
spectra are affected by resonance decays). At RHIC there has been speculation
that the baryon to meson anomaly (and elliptic flow) observed at intermediate
\pt, $2 < \pt < \gevc{8}$, is related to the recombination of flowing valence
quark like degrees of freedom rather than hydrodynamic flow. There have been
predictions that these effects should extend out to much higher \pt at
LHC~\cite{Hwa:2006zq}. At high \pt, $\pt > \gevc{8}$, the observed PID effects
should be mainly due to the interaction of the hard probe with the
medium. Following~\cite{Sapeta:2007ad} we can imagine that the hard parton
directly exchanges quantum numbers, e.g. baryon number, with the medium, but
also that the color flow of the fragmentation is modified by radiative energy
loss and medium partons, and that this gives large effects due to the changes
in invariant mass. The latter effect seems very generic, and
in~\cite{Sapeta:2007ad} this interplay is modeled via enhanced parton
splitting functions and they find large effects on the particle ratios
(relative to ``\pp fragmentation'') inside the jets even out to very large
\pt. It is in particular these high \pt PID effects we are interested in
addressing here.

\section{High \pt PID in the ALICE experiment}

\begin{figure}[htbp]
  \begin{center}
    \includegraphics[keepaspectratio, width=0.5\columnwidth]{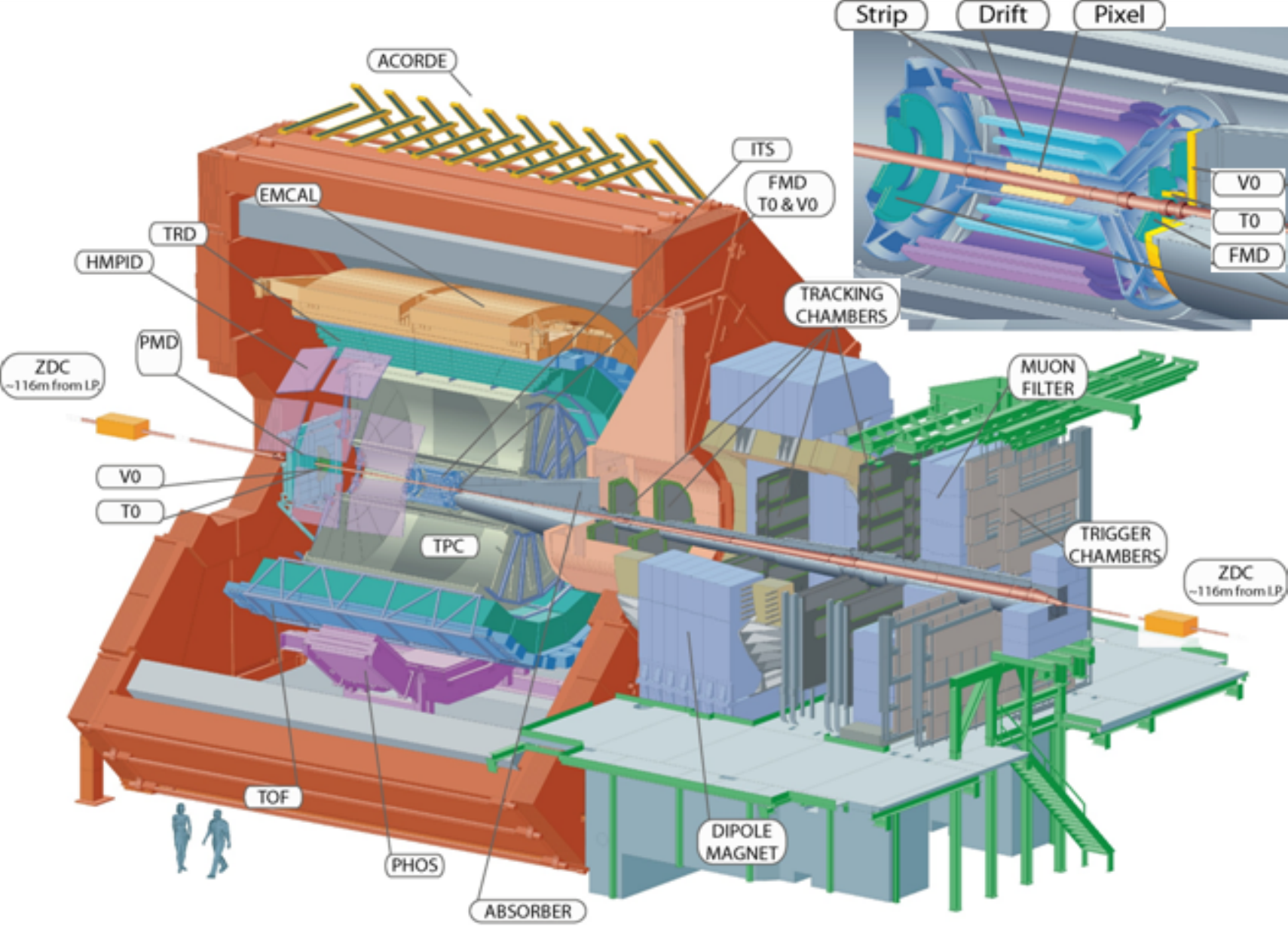}

  \end{center}
  \caption{Schematic view of the ALICE experiment. The main detector used for
    the analysis reported here is the TPC located near the center of the the
    central barrel (inside the L3 magnet).}
  \label{fig:alice}
\end{figure}

Figure~\ref{fig:alice} shows the layout of the ALICE
experiment~\cite{Aamodt:2008zz}. ALICE is a dedicated heavy ion experiment
with full azimuthal coverage around mid-rapidity (central barrel located
inside the L3 magnet) and a dedicated forward muon tracking system. The
results reported here rely mainly on the excellent tracking and PID
capabilities of the Time Projection Chamber (TPC)~\cite{Alme:2010ke}. The \pt
resolution for primary tracks associated with hits in the Silicon based Inner
Tracking System (ITS) is better than 5~\% at $\pt = \gevc{20}$.

In the ALICE experiment it is possible to identify particles with very high
transverse momentum, $\pt \gg \gevc{3}$. Charged pions and kaons+protons
(together) can be identified from the \dedx, thanks to the separation on the
relativistic rise, and \KOs and \LA can be identified from their \VZ weak
decay topology~\cite{Belikov:2011zz}. The identification of $\uppi^0$ with the
calorimeters and via conversion of photons was covered in another
presentation~\cite{Yury}.

\section{High \pt results}

ALICE has recently submitted results on identified flow, $v_2$ and $v_3$, at
high \pt for publication~\cite{Abelev:2012di}. The main results we shall quote
from there is that for $\pt > \gevc{8}$, $v_3$ is small and there does not
seem to be large PID effects for $v_2$. This suggests that at high \pt genuine
flow effects are small, i.e., we are in a dominantly hard/jet regime.

In the rest of this section we discuss PID on the relativistic rise of the TPC
\dedx. The \dedx is obtained as the truncated mean of the $0-60\%$ lowest
charge samples. The performance and stability, with respect to
e.g. pressure variations, of the \dedx is improved in the following way: space
points that only deposits charge on 1 pad, which are not used for track
fitting, are included, and missing hits in between rows where hits are found
are assigned a virtual charge of the lowest reconstructed charge cluster on
the track to account for threshold effects.

\begin{figure}[htbp]
  \begin{center}
    \includegraphics[keepaspectratio, width=0.5\columnwidth]{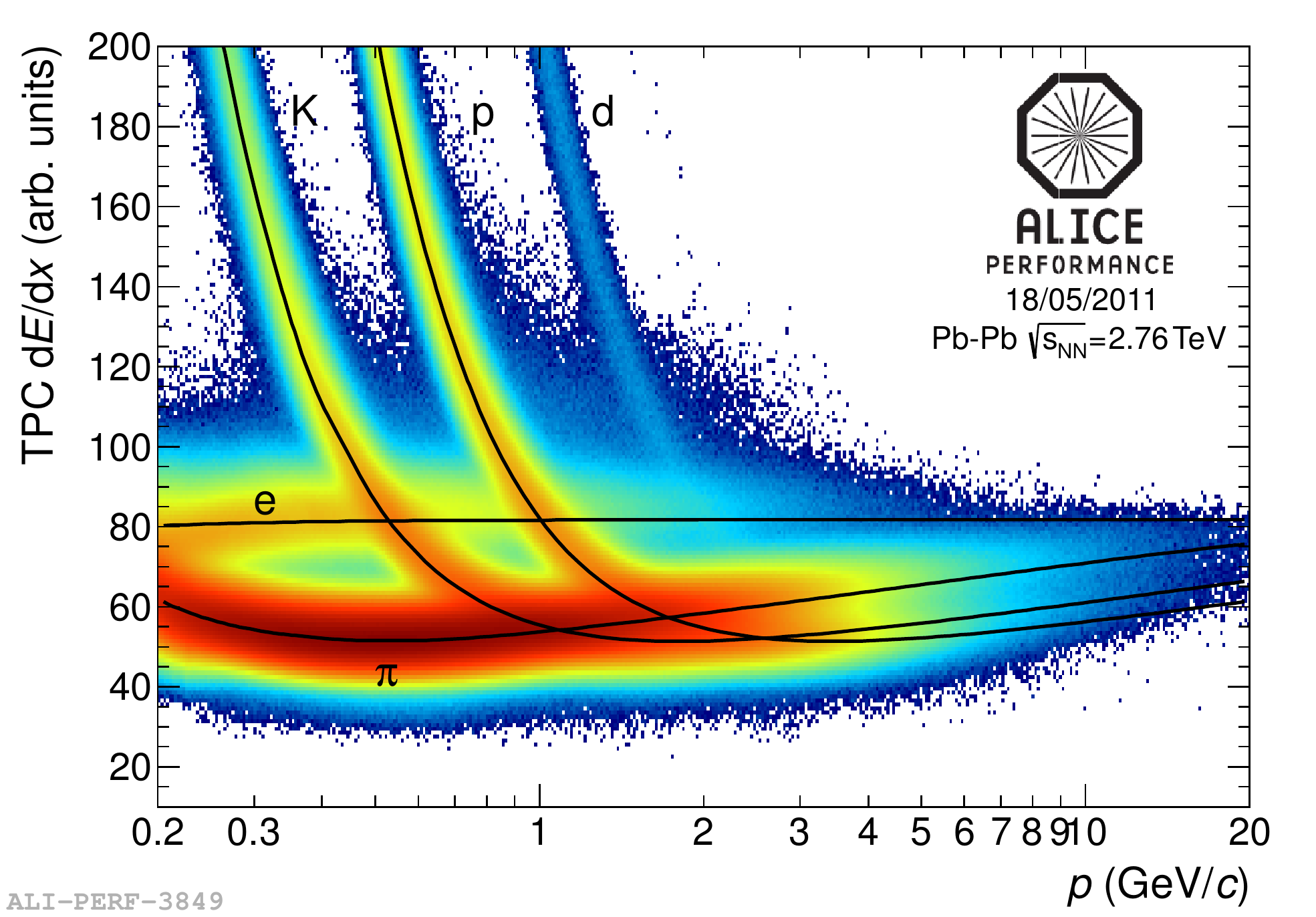}
  \end{center}
  \caption{TPC \dedx vs $p$. The curves show the \mdedx for $\uppi$, K, p, and e.}
  \label{fig:tpc_dedx}
\end{figure}
 
Figure~\ref{fig:tpc_dedx} shows the \dedx vs $p$ for \pbpb data. We note that
for $p>\gevc{3}$ pions, kaons, and protons can in principle be separated. The
first step in the analysis is the extraction of parameterizations for
$\mdedx(\beta\gamma)$ and $\sigma(\mdedx)$. The extraction is done
independently for each \pp sample and centrality class using a 2-dimensional
fit to similar data as shown in the figure. For this analysis a constant
relative resolution, $\sigma/\mdedx = const$, has been used.

\begin{figure}[htbp]
  \begin{center}
    \includegraphics[keepaspectratio, width=0.31\columnwidth]{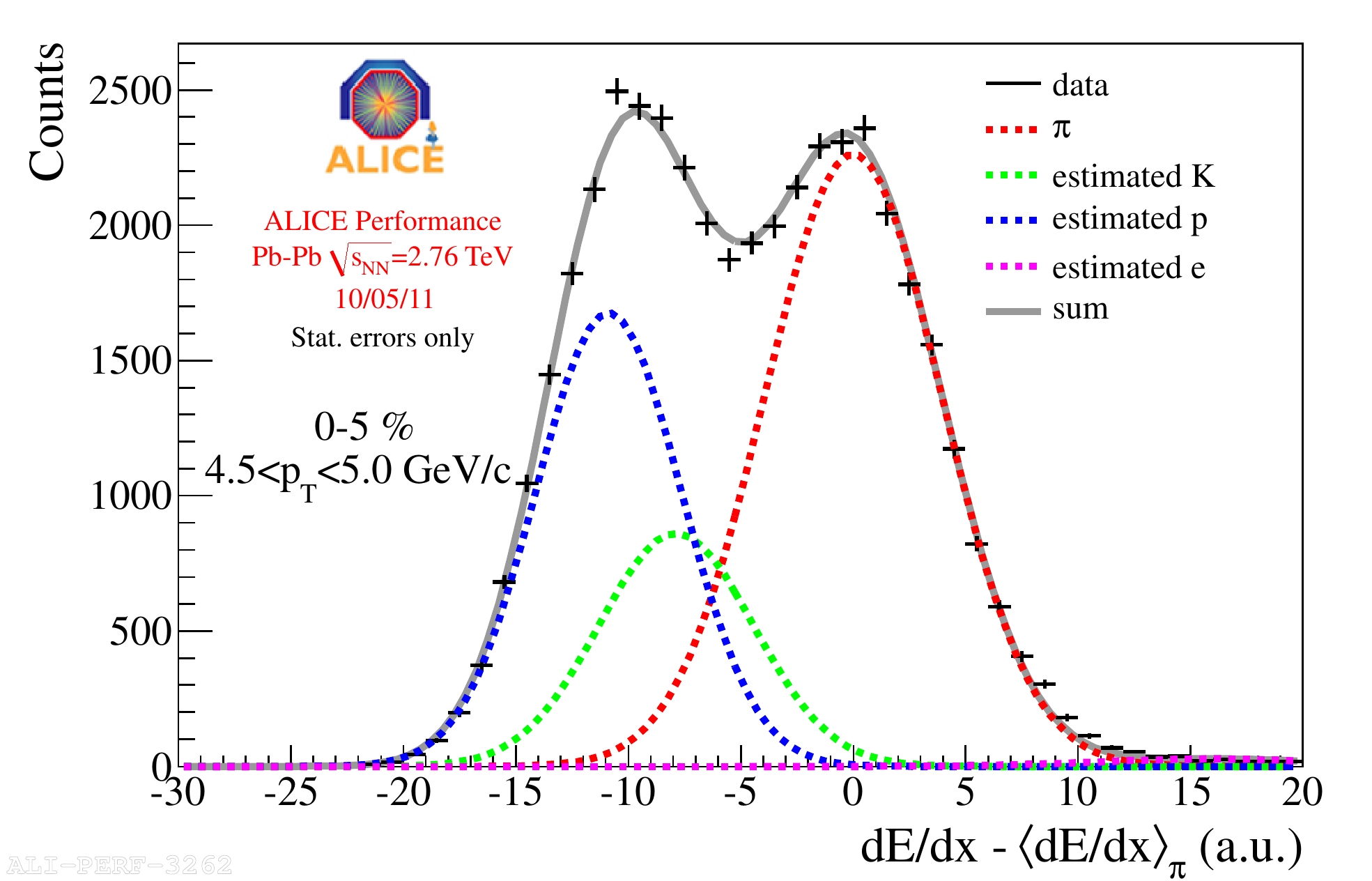}
    \includegraphics[keepaspectratio, width=0.31\columnwidth]{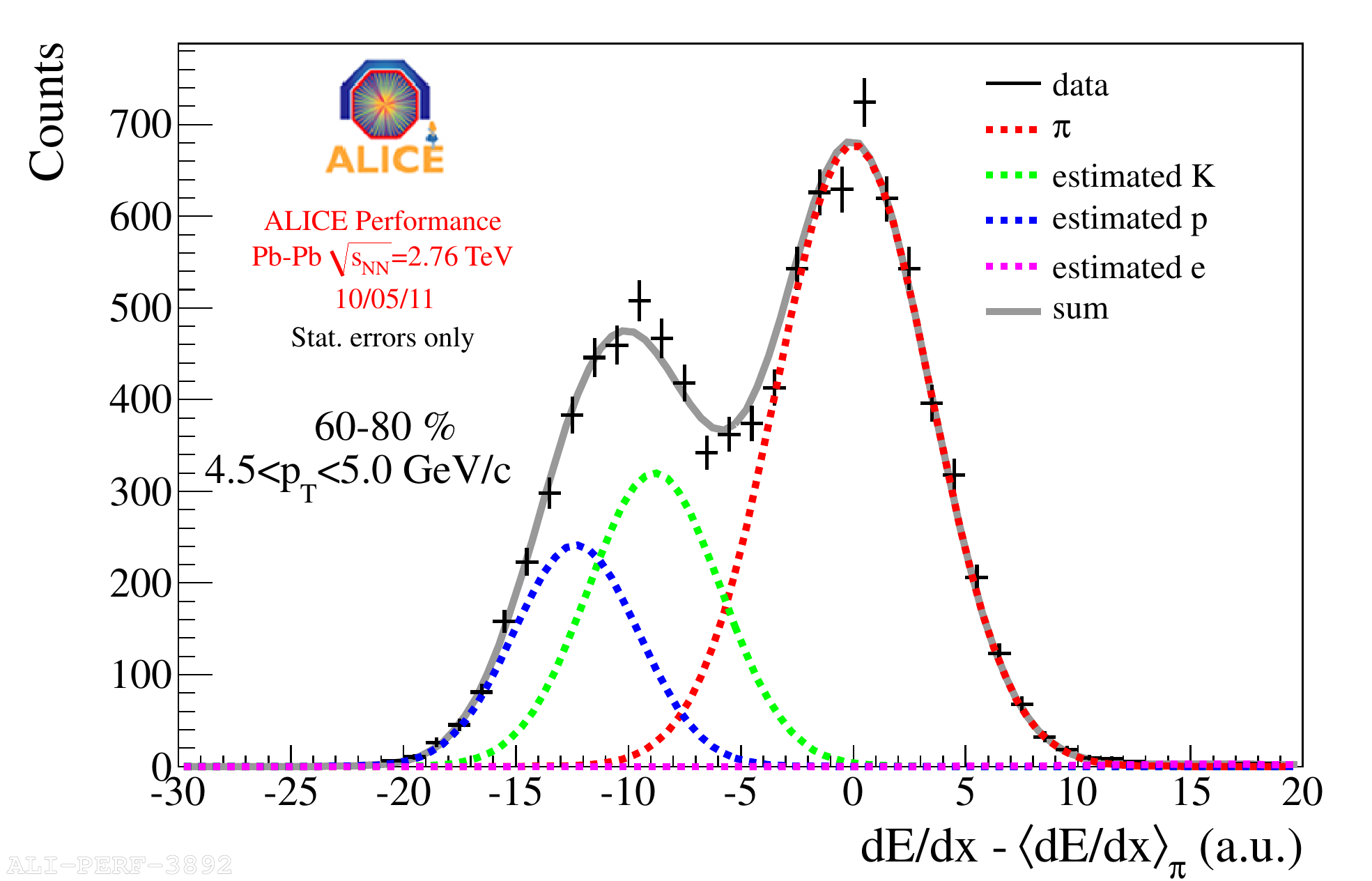}
    \includegraphics[keepaspectratio, width=0.31\columnwidth]{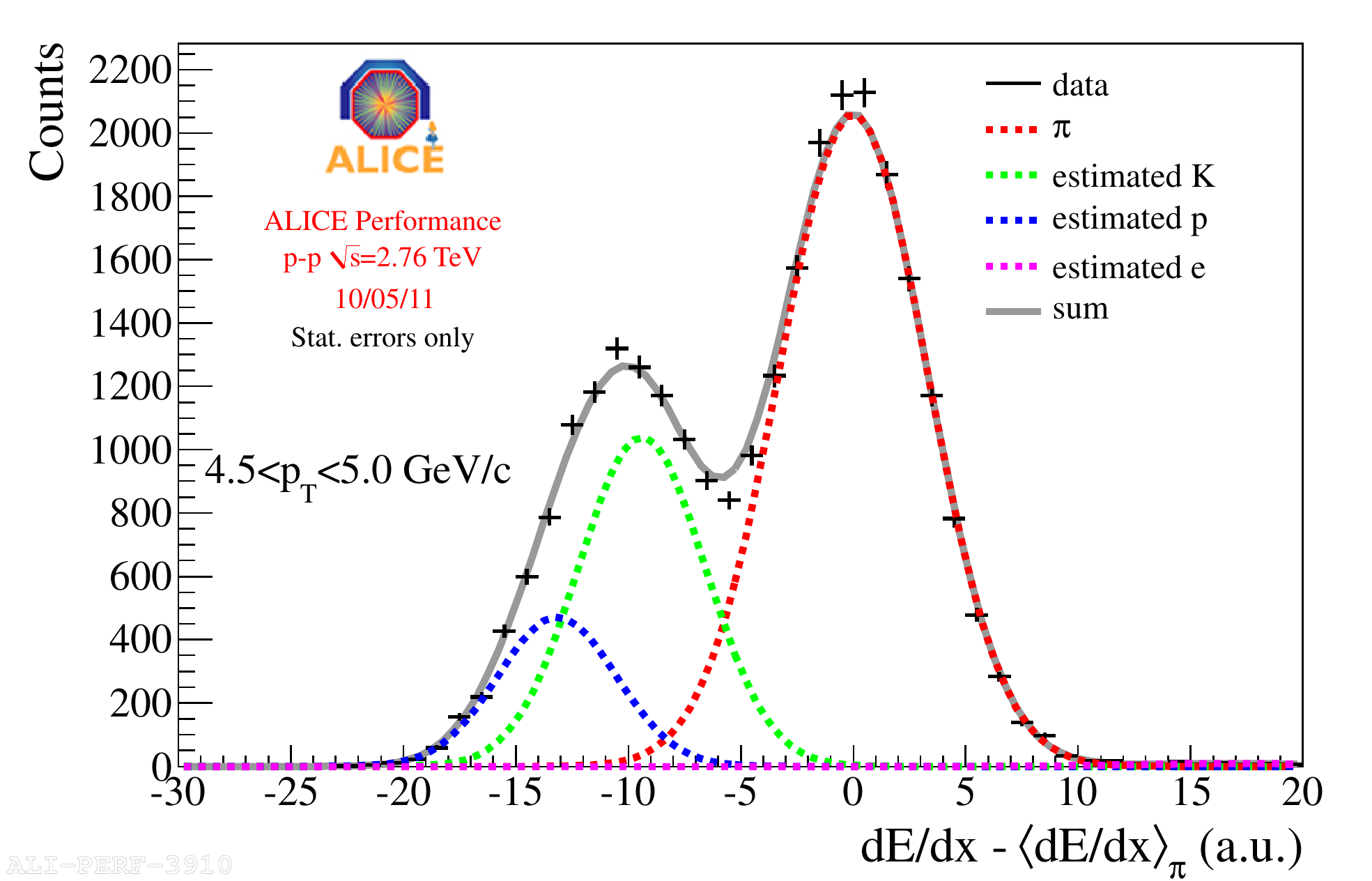}
  \end{center}
  \begin{center}
    \includegraphics[keepaspectratio, width=0.31\columnwidth]{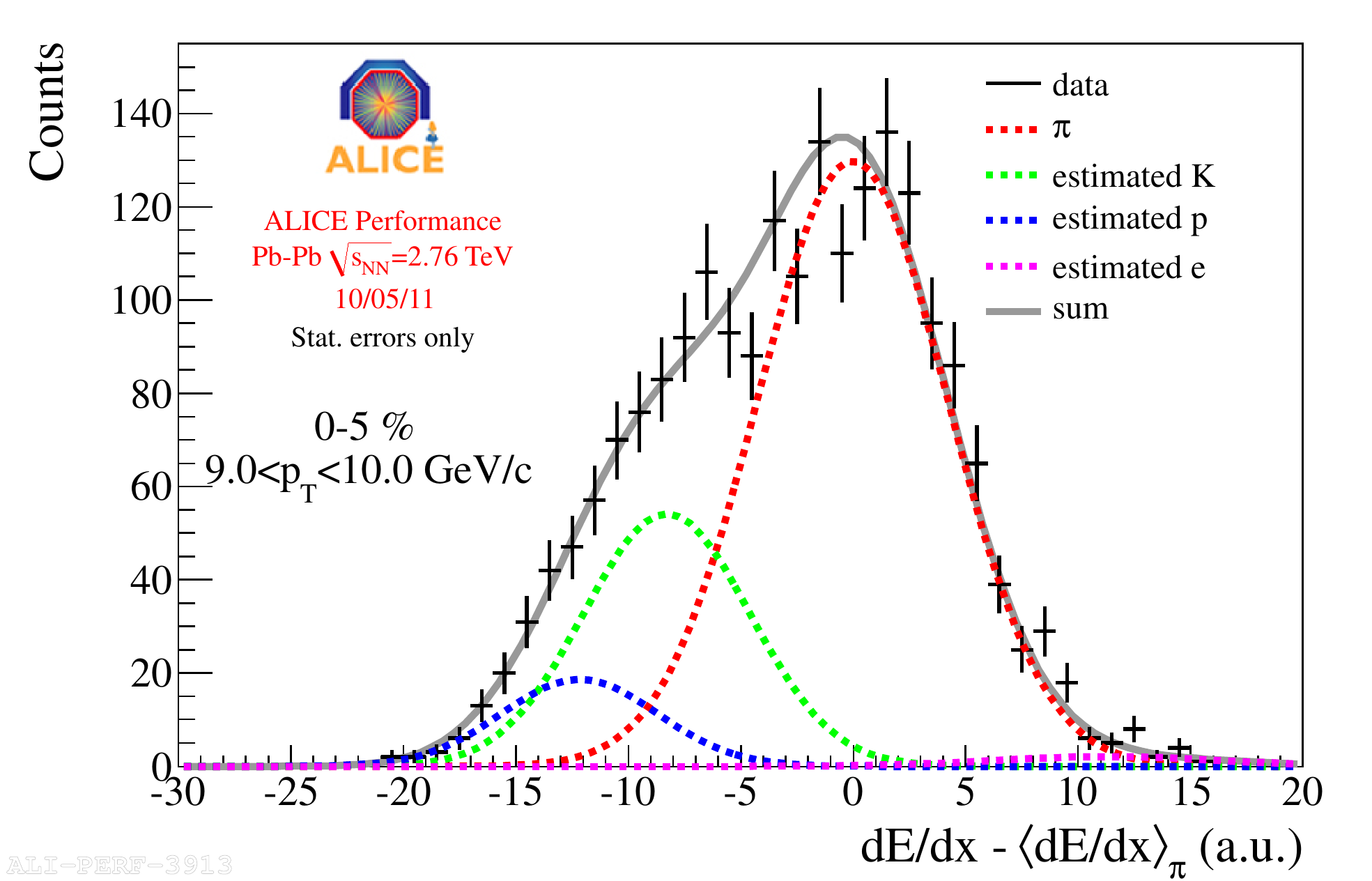}
    \includegraphics[keepaspectratio, width=0.31\columnwidth]{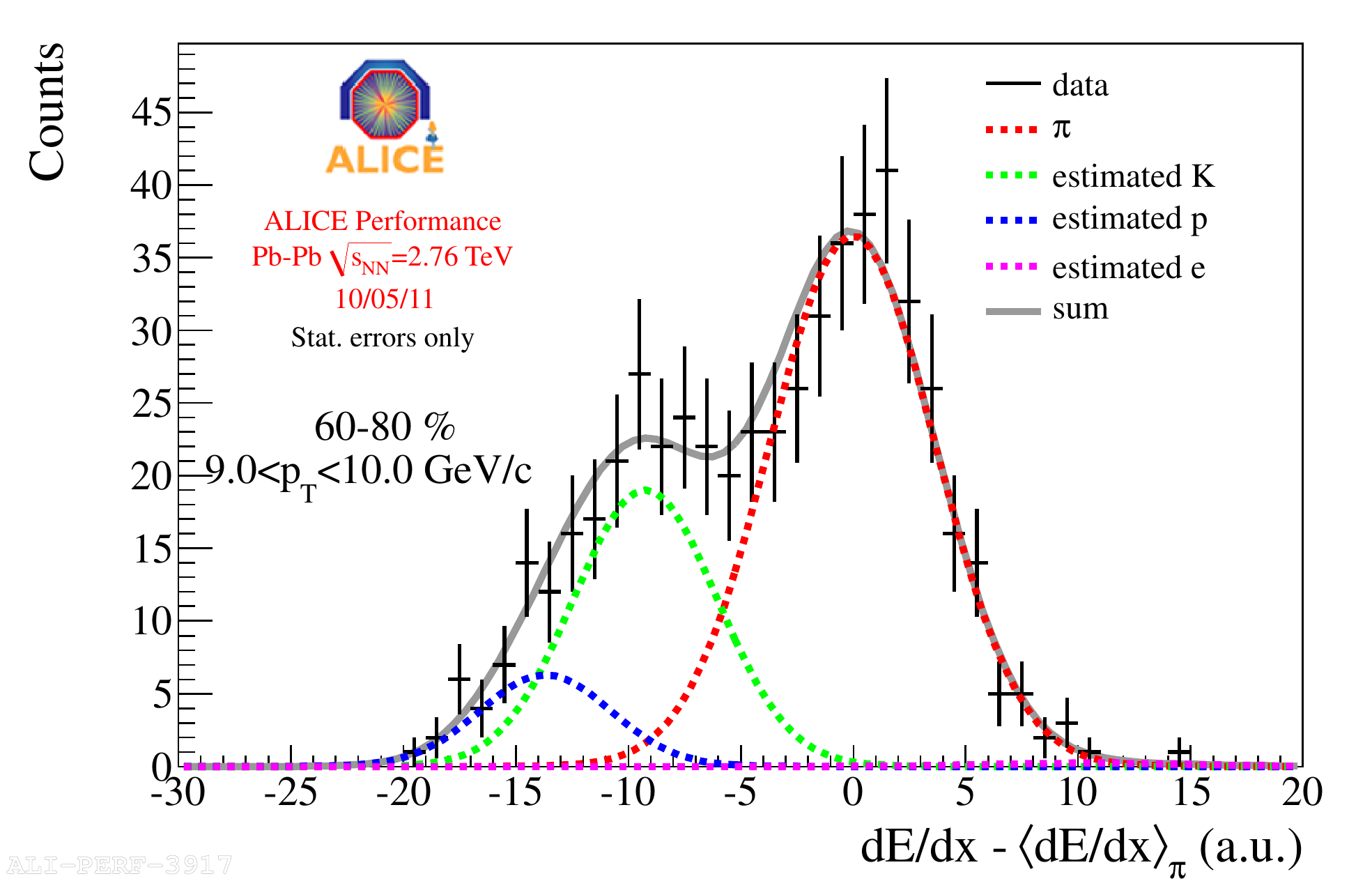}
    \includegraphics[keepaspectratio, width=0.31\columnwidth]{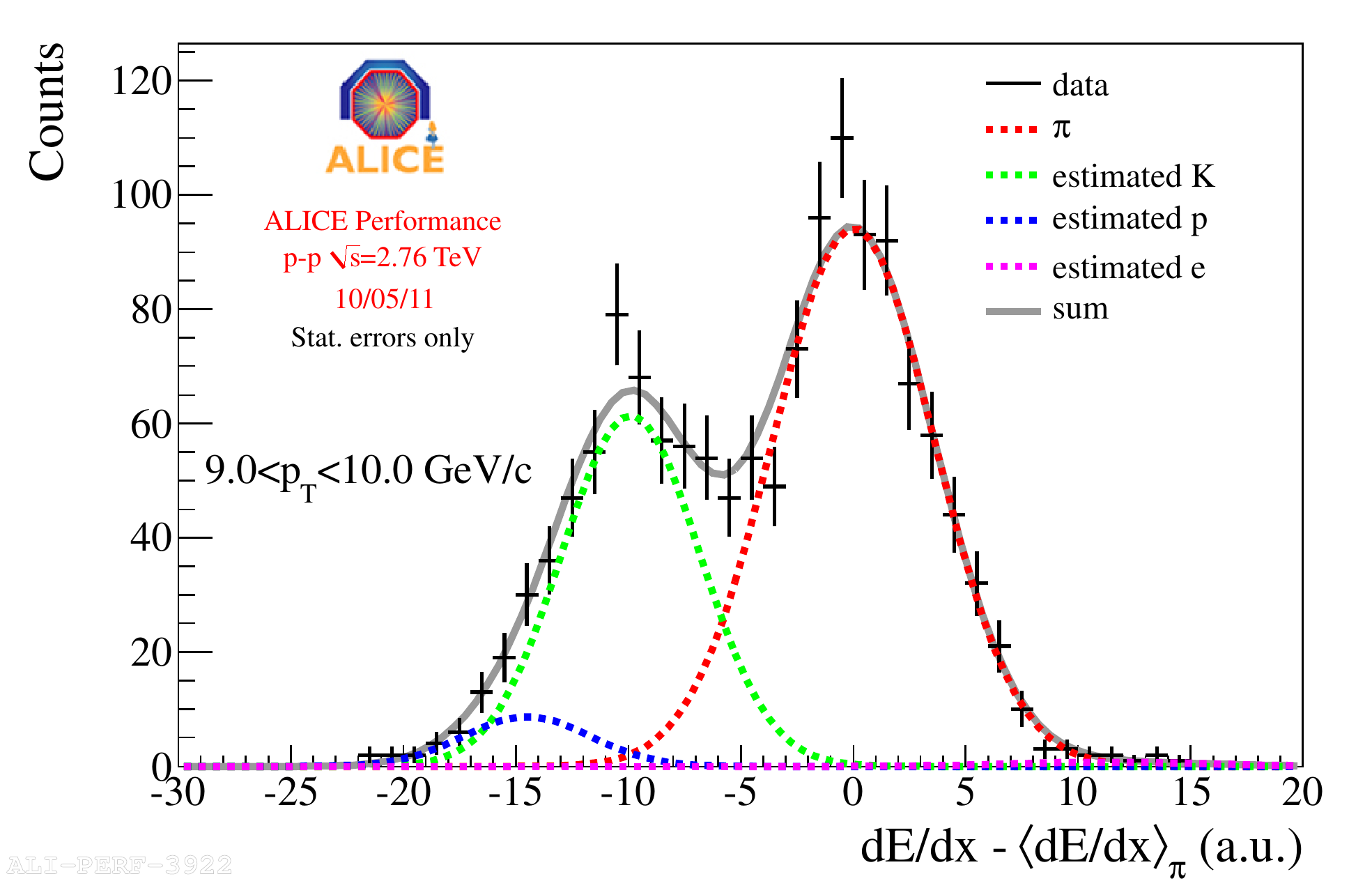}
  \end{center}
  \caption{\dpi distributions fitted with a sum of 4 Gaussians for two \pt
    intervals, $4.5 < \pt < \gevc{5.0}$ (upper) and $9.0 < \pt < \gevc{10.0}$
    (lower), in central (left) and peripheral (center) \pbpb, and \pp (right)
    collisions.}
  \label{fig:prelim:dpi_fits}
\end{figure}

For PID the quantity $\dpi = \dedx - \mdedxpi$ has been studied as a function
of \pt. Figure~\ref{fig:prelim:dpi_fits} shows an example of \dpi spectra for
different data sets in 2 \pt intervals. The estimated distributions are fitted
using a sum of 4 Gaussians ($\uppi$, K, p, and e) where the mean and width of
each Gaussian has been constrained from the parameterizations of \mdedx and
$\sigma$. It is clear already from the \dpi spectra that the composition of
particle species is very different in central \pbpb from peripheral \pbpb and
\pp. Furthermore this difference seems to be greatly reduced or gone at higher
\pt. This is similar to the baryon-meson enhancement observed for
$\LA/\KOs$~\cite{Belikov:2011zz}, but we do not here try to separate the kaons
and protons in the \dpi distributions (this analysis was shown at Quark Matter
2012 and needed a refined description of \mdedx and $\sigma$).

From the fit to the data we extract the fraction of pions. To extract pion
spectra we use the \dndeta{\ch} of unidentified charged
particles~\cite{Michele} to normalize the results using the equation:
\begin{equation}
  \dndeta{\pi} = \dndeta{\ch} \times \frac{\eff{\ch}}{\eff{\pi}} \times \frac{\yield{\pi}}{\yield{\ch}},
\end{equation}
where $\yield{\pi}/\yield{\ch}$ is the uncorrected pion fraction obtained from
fits like in Figure~\ref{fig:prelim:dpi_fits} and $\eff{\pi}/\eff{\ch}$ is the
relative pion efficiency which is independent, within a 2~\% systematic
uncertainty, of centrality and \pt in the measured interval. To obtain
rapidity spectra, a small correction is applied to convert the pseudorapidity
interval ($|\eta| < 0.8$) into a rapidity interval.

The dominating systematic error on the extracted pion fraction has been
estimated by releasing the constraints used in the \dpi fits. It is around
3~\% for \pp and ~5\% for \pbpb. For the spectra and \RAA, the full systematic
error of the unidentified analysis is also taken over~\cite{Michele}.

\begin{figure}[htbp]
  \centering
  \includegraphics[keepaspectratio, width=0.4\columnwidth]{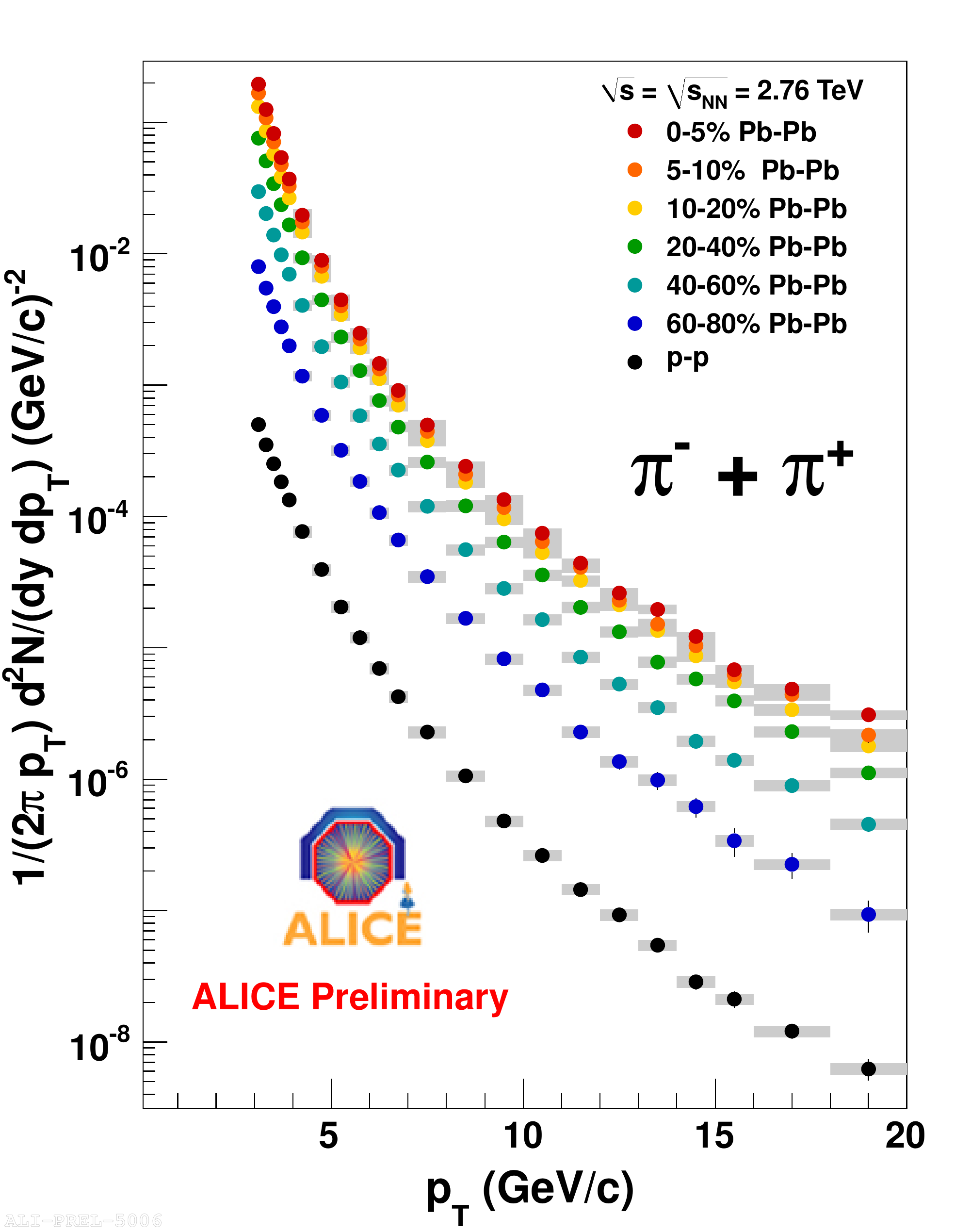}
  \includegraphics[keepaspectratio, width=0.4\columnwidth]{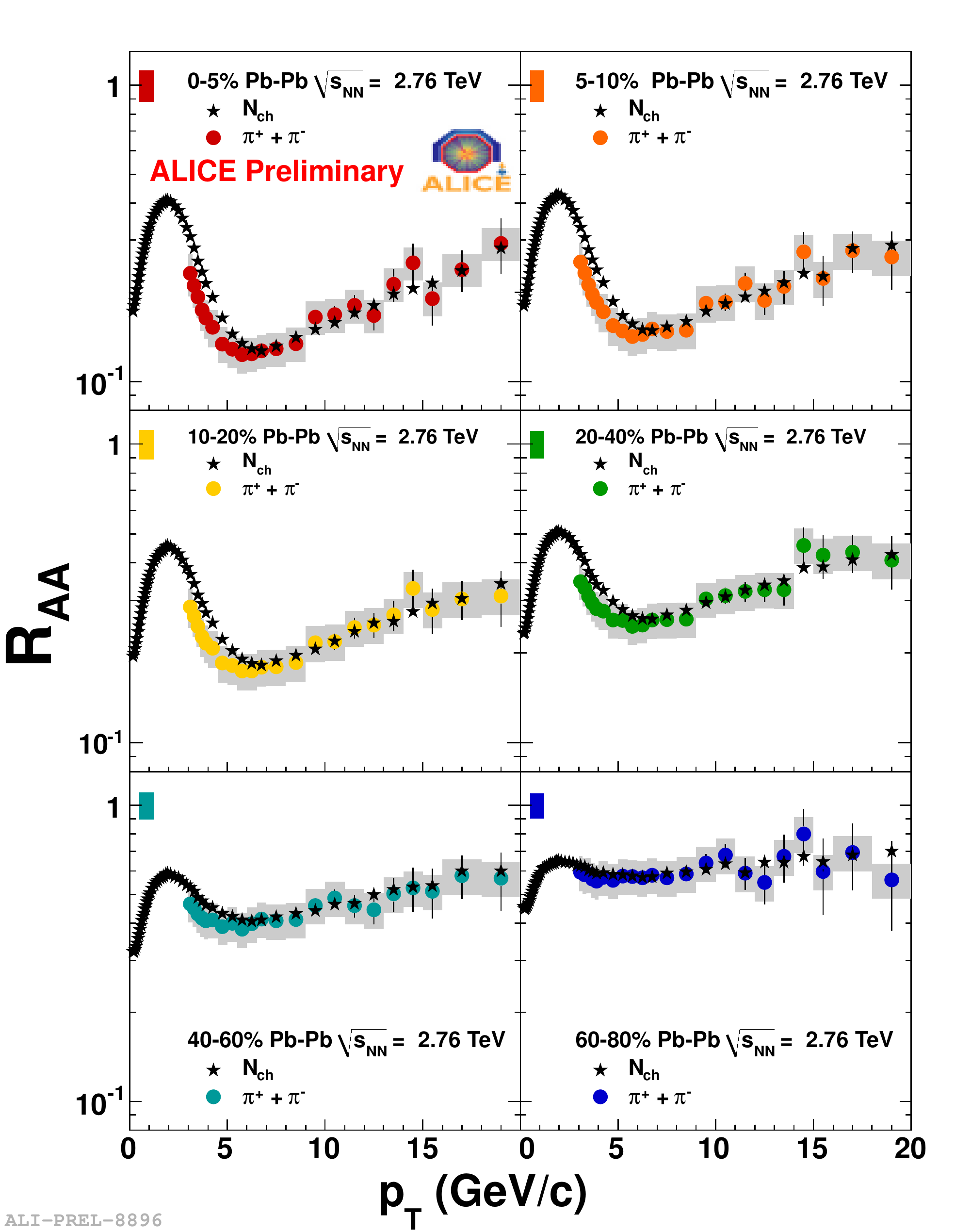}
  \caption{Left: the invariant yield for charged pions, \sumpi, as a function
    of \pt for different \pbpb centrality classes and \pp. Statistical errors
    are shown by the vertical error bars. Systematic errors are shown as gray
    boxes. Right: the \RAA computed from these spectra and compared to the
    \RAA for unidentified charged particles (black points) as a function of
    \pt for different centrality classes. Statistical (vertical error bars)
    and systematic (gray and colored boxes) are shown for the charged \sumpi
    \RAA. The colored boxes contains the common systematic error related to
    the nuclear overlap function and the \pp normalization to the total
    inelastic cross section. Only
    statistical errors are shown for the unidentified charged \RAA.}
  \label{fig:prelim:pion_spectra}
\end{figure}

Figure~\ref{fig:prelim:pion_spectra} (left) shows the spectra for charged pions,
\sumpi, for $3 < \pt < \gevc{20}$. From these spectra the \RAA can be
computed:

\begin{equation}
  \RAA = \frac{\dndypbpb}{\langle \TAA \rangle \dndsigmapp},
\end{equation}
where $\langle \TAA \rangle$ is the nuclear overlap function obtained from
a Glauber calculation for a given centrality class.

Figure~\ref{fig:prelim:pion_spectra} (right) shows the \RAA for charged
pions. For $\pt < \gevc{8}$ charged pions are more suppressed than bulk
unidentified particles, while for $\pt > \gevc{8}$ the suppression is
similar. 

\begin{figure}[htbp]
  \centering
  \includegraphics[keepaspectratio, width=0.85\columnwidth]{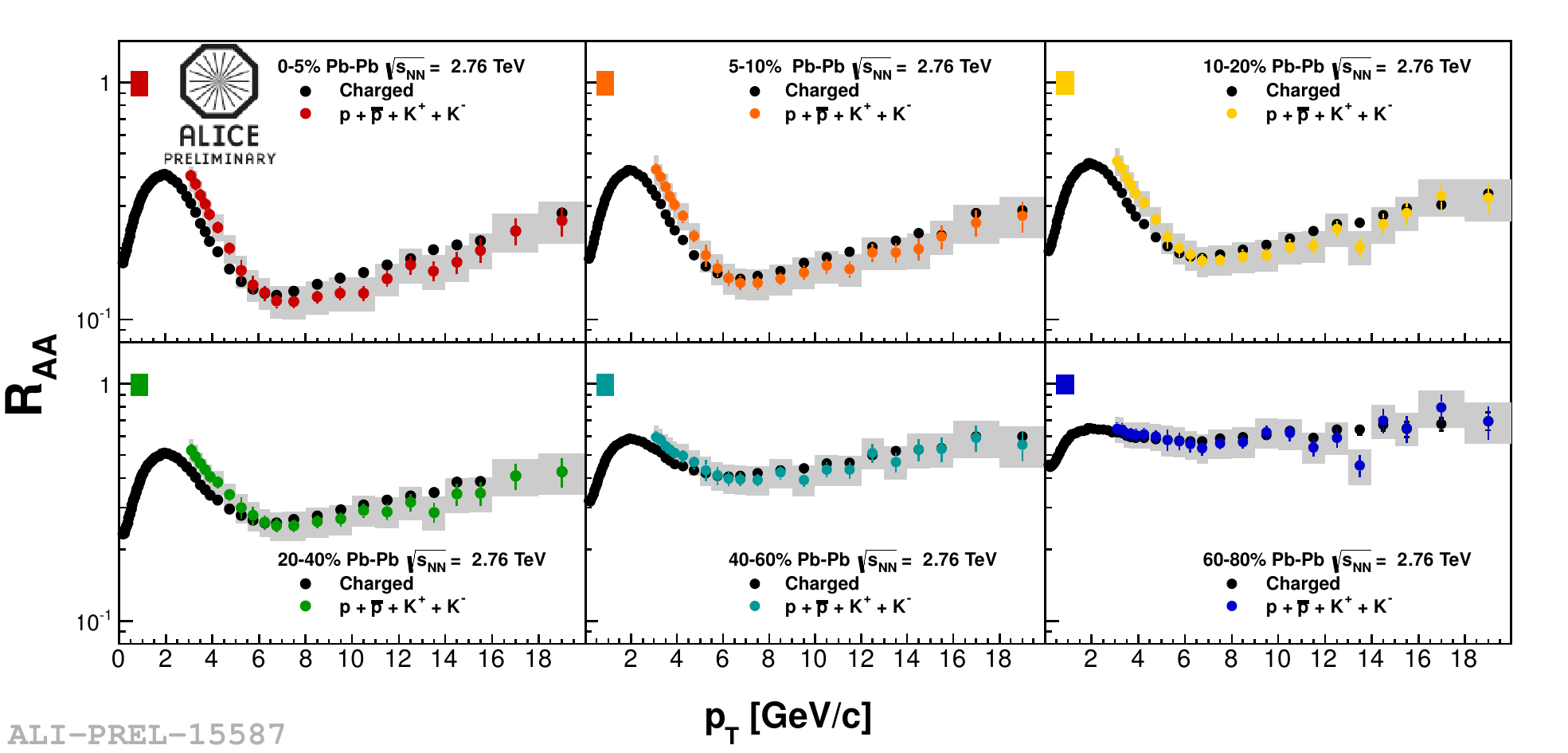}
  \caption{The figure shows the \RAA for the sum of kaons and protons compared
    to the \RAA for unidentified charged particles as a function of \pt for
    different centrality classes. Statistical (vertical error bars) and
    systematic (gray and colored boxes) are shown for the charged K+p
    \RAA. The colored boxes contains the common systematic error related to
    the nuclear overlap function and the \pp normalization to the total
    inelastic cross section. Only statistical errors are shown for the
    unidentified charged \RAA.}
  \label{fig:res:raa_kp}
\end{figure}

Even we yet do not trust separately the fits of protons and kaons, the sum is
stable. To enhance the significance of the previous results we can therefore
make a similar analysis for the sum of K+p (\sumkp). The absolute magnitude of
the systematic error is similar to that of pions, and the variation of the
yield due to slightly different efficiency for K and p and rapidity correction
even when changing the lower yield by $\pm 50~\%$ is much smaller.

Figure~\ref{fig:res:raa_kp} shows the results for \RAA. For $\pt < \gevc{8}$
charged K+p is less suppressed than bulk unidentified particles, as expected
since pions are more suppressed in this region. For $\pt > \gevc{8}$ the
suppression is similar.

\begin{figure}[htbp]
  \centering
  \includegraphics[keepaspectratio, height=0.35\columnwidth]{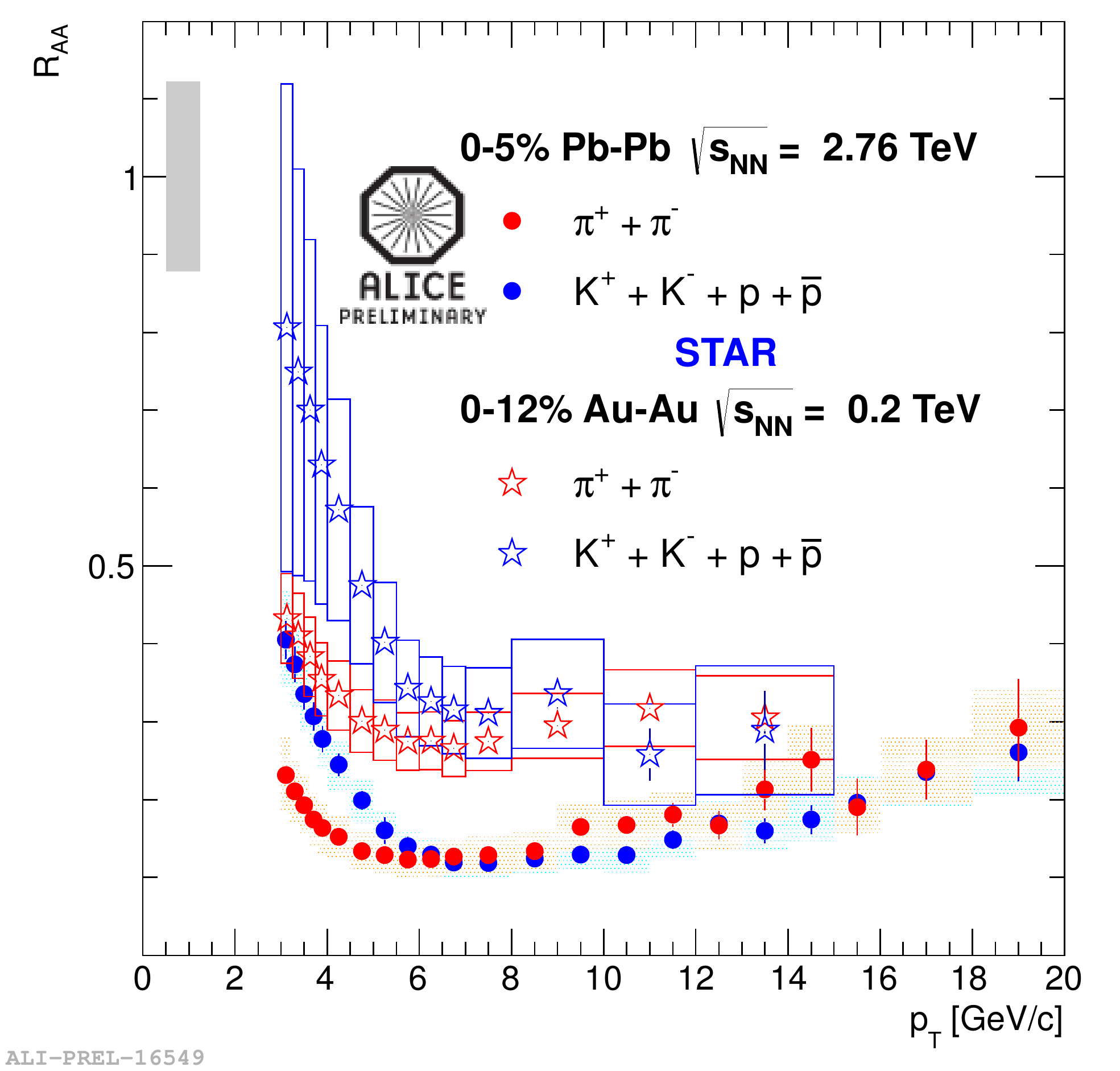}
  \includegraphics[keepaspectratio, height=0.35\columnwidth]{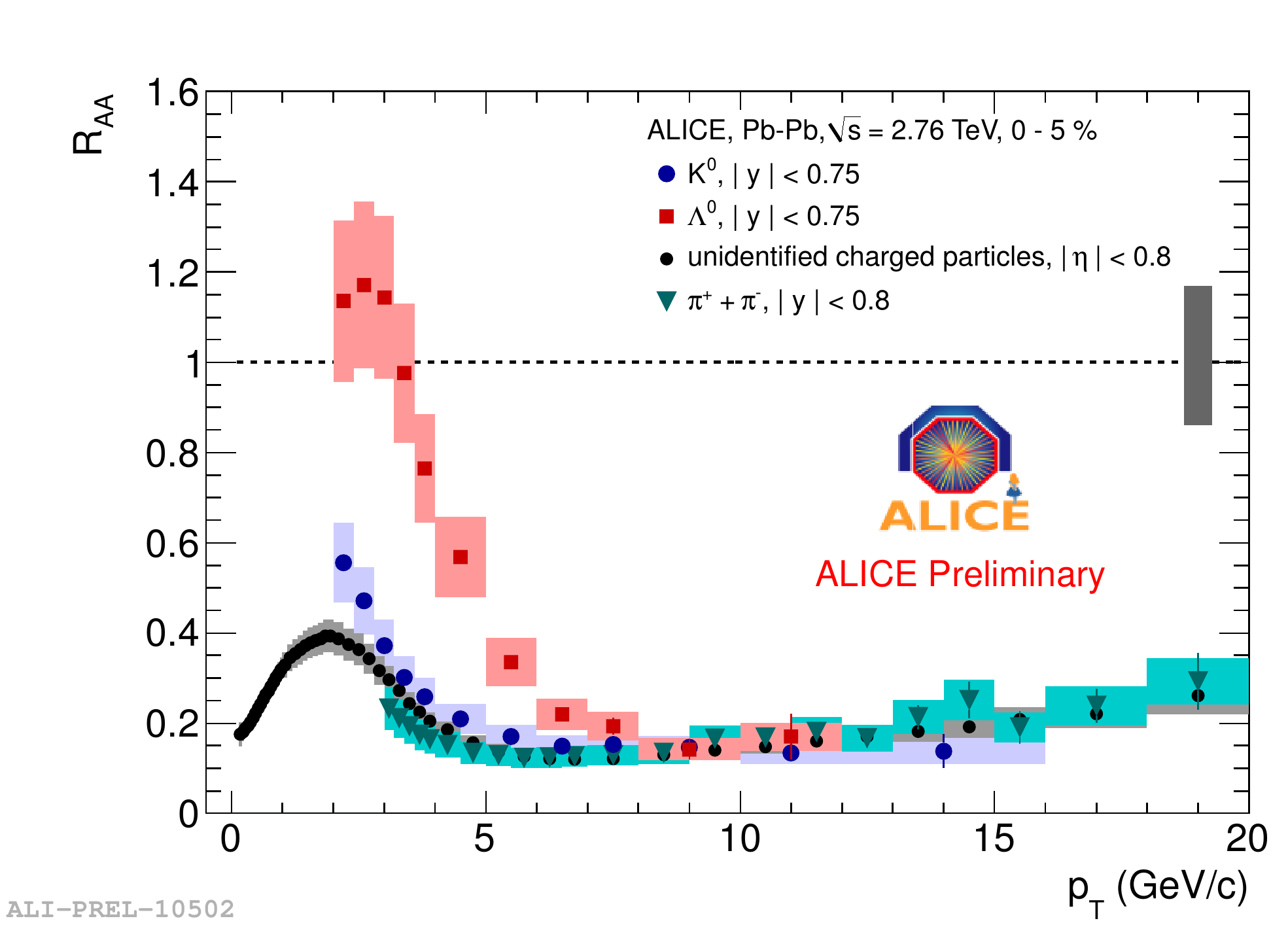}
  \caption{Left: \RAA for the sum of kaons and protons compared
    to the \RAA for pions for \pbpb 0-5~\%. Similar measurements by the STAR
    experiment~\cite{Agakishiev:2011dc} are also included. Right: \RAA summary
    for light quark hadrons for \pbpb 0-5~\%.}
  \label{fig:res:raa_all}
\end{figure}

Figure~\ref{fig:res:raa_all} (left) shows a comparison between the results for
charged pions and the sum of kaons and protons for the 0-5~\% most central
collisions. Similar results from STAR at RHIC have also been included. The
result indicates that at high \pt large differences in suppression is only
possible for K and p separately. This is contrary to~\cite{Sapeta:2007ad}
where the particle ratios for $\text{K}/\uppi$ and $\text{p}/\uppi$ were both
found to be enhanced within jets (and so $\text{K+p}/\uppi$ in central
collisions is much larger than in \pp collisions)~\footnote{Recall that
  mathematically $\RAA(\ssumkp)/\RAA(\ssumpi) =
  ((\ssumkp)/\ssumpi)_{\text{\pbpb}}/((\ssumkp)/\ssumpi)_{\text{\pp}}$.}.

\section{Discussion}

Figure~\ref{fig:res:raa_all} (right) shows a summary of \RAA for inclusive
charged particles and some light quark hadrons: \sumpi, \KOs and \LA. The
preliminary results from ALICE on \RAA for identified particles all suggest
that at high \pt light quark hadrons are equally suppressed. The question
which I want to speculate on here is what this results could indicate in terms
of quenching.

In general for \RAA of high \pt particles we are sensitive mostly to leading
particle effects and there is some bias towards the surface and unmodified
jets. However, ALICE also presented results at Hard Probes 2012 that shows
that the p$/\uppi$ ratio in a region around high \pt triggers, after the bulk
contribution has been subtracted, is similar to the expectations for
\pp~\cite{Veldhoen}. STAR also presented a similar study but with less clear
conclusions~\cite{Davila}. The bulk ratios can of course be affected by
radiative energy loss and one would benefit from more advanced methods, but
the results suggest that identified particle production of subleading
particles in jets is also not strongly modified.

If one compares to the possible PID effects at high \pt mentioned in the
introduction, the most generic process that seems to be ruled out is strong
modified color (gluon) flow in the fragmentation
process~\cite{Sapeta:2007ad}. It seems natural to assume that once the parton
starts propagating though the medium both radiative energyloss and shower
radiation occurs. Taken these results to the extreme one might then propose
that the parton fragmentation differentiates between these two types of
radiation. One would then have to establish the physics that would decouple
these processes, e.g., the energy loss could be radiated as ``free'' gluons.

There is a large theoretical activity on quantum interference effects and the
effect on the energy loss/fragmentation. These results do not directly support
the picture above, but have some of the ingredients in terms of coherence and
decoherence effects that are different in the medium and in the vacuum, see
e.g.~\cite{Tywunik} and references therein.

It is clear that the results from ALICE reported here could play an important
role in guiding these challenging theoretical efforts.

\section{Conclusions}

The \RAA for $\uppi$, K+p, $\LA$, and \KOs at high \pt ($\pt \gg \gevc{8}$)
seems to indicate that particle species dependent effects are, if present,
small. This provides important input to models of the energy loss and may
restrict color flow effects.





\bibliographystyle{elsarticle-num}
\bibliography{proceeding.bib}

\begin{thebibliography}{10}
\expandafter\ifx\csname url\endcsname\relax
  \def\url#1{\texttt{#1}}\fi
\expandafter\ifx\csname urlprefix\endcsname\relax\def\urlprefix{URL }\fi
\expandafter\ifx\csname href\endcsname\relax
  \def\href#1#2{#2} \def\path#1{#1}\fi

\bibitem{Aamodt:2010jd}
K.~Aamodt, et~al., {Suppression of Charged Particle Production at Large
  Transverse Momentum in Central Pb--Pb Collisions at $\sqrt{s_{NN}} = 2.76$
  TeV}, Phys.Lett. B696 (2011) 30--39.
\newblock \href {http://arxiv.org/abs/1012.1004} {\path{arXiv:1012.1004}},
  \href {http://dx.doi.org/10.1016/j.physletb.2010.12.020}
  {\path{doi:10.1016/j.physletb.2010.12.020}}.

\bibitem{Adcox:2003nr}
K.~Adcox, et~al., {Single identified hadron spectra from $\sqrt{s_{NN}}$ =
  130-GeV Au+Au collisions}, Phys.Rev. C69 (2004) 024904.
\newblock \href {http://arxiv.org/abs/nucl-ex/0307010}
  {\path{arXiv:nucl-ex/0307010}}, \href
  {http://dx.doi.org/10.1103/PhysRevC.69.024904}
  {\path{doi:10.1103/PhysRevC.69.024904}}.

\bibitem{Adams:2006wk}
J.~Adams, et~al., {Measurements of identified particles at intermediate
  transverse momentum in the STAR experiment from Au + Au collisions at
  $\sqrt{s_{NN}}$ =200- GeV}\href {http://arxiv.org/abs/nucl-ex/0601042}
  {\path{arXiv:nucl-ex/0601042}}.

\bibitem{Abelev:2006jr}
B.~Abelev, et~al., {Identified baryon and meson distributions at large
  transverse momenta from Au+Au collisions at $\sqrt{s_{NN}}$ = 200-GeV},
  Phys.Rev.Lett. 97 (2006) 152301.
\newblock \href {http://arxiv.org/abs/nucl-ex/0606003}
  {\path{arXiv:nucl-ex/0606003}}, \href
  {http://dx.doi.org/10.1103/PhysRevLett.97.152301}
  {\path{doi:10.1103/PhysRevLett.97.152301}}.

\bibitem{Hwa:2006zq}
R.~C. Hwa, C.~Yang, {Proton enhancement at large p(T) at LHC without structure
  in associated-particle distribution}, Phys.Rev.Lett. 97 (2006) 042301.
\newblock \href {http://arxiv.org/abs/nucl-th/0603053}
  {\path{arXiv:nucl-th/0603053}}, \href
  {http://dx.doi.org/10.1103/PhysRevLett.97.042301}
  {\path{doi:10.1103/PhysRevLett.97.042301}}.

\bibitem{Sapeta:2007ad}
S.~Sapeta, U.~A. Wiedemann, {Jet hadrochemistry as a characteristics of jet
  quenching}, Eur.Phys.J. C55 (2008) 293--302.
\newblock \href {http://arxiv.org/abs/0707.3494} {\path{arXiv:0707.3494}},
  \href {http://dx.doi.org/10.1140/epjc/s10052-008-0592-8}
  {\path{doi:10.1140/epjc/s10052-008-0592-8}}.

\bibitem{Aamodt:2008zz}
K.~Aamodt, et~al., {The ALICE experiment at the CERN LHC}, JINST 3 (2008)
  S08002.
\newblock \href {http://dx.doi.org/10.1088/1748-0221/3/08/S08002}
  {\path{doi:10.1088/1748-0221/3/08/S08002}}.

\bibitem{Alme:2010ke}
J.~Alme, Y.~Andres, H.~Appelshauser, S.~Bablok, N.~Bialas, et~al., {The ALICE
  TPC, a large 3-dimensional tracking device with fast readout for ultra-high
  multiplicity events}, Nucl.Instrum.Meth. A622 (2010) 316--367.
\newblock \href {http://arxiv.org/abs/1001.1950} {\path{arXiv:1001.1950}},
  \href {http://dx.doi.org/10.1016/j.nima.2010.04.042}
  {\path{doi:10.1016/j.nima.2010.04.042}}.

\bibitem{Belikov:2011zz}
I.~Belikov, {K0(S) and Lambda production in Pb-Pb collisions with the ALICE
  experiment}, J.Phys.G G38 (2011) 124078.
\newblock \href {http://dx.doi.org/10.1088/0954-3899/38/12/124078}
  {\path{doi:10.1088/0954-3899/38/12/124078}}.

\bibitem{Yury}
Y.~Karlov, these proceedings (hard probes 2012).

\bibitem{Abelev:2012di}
B.~Abelev, et~al., {Anisotropic flow of charged hadrons, pions and
  (anti-)protons measured at high transverse momentum in Pb-Pb collisions at
  $\sqrt{s_{NN}}=2.76$ TeV}\href {http://arxiv.org/abs/1205.5761}
  {\path{arXiv:1205.5761}}.

\bibitem{Michele}
M.~Floris, these proceedings (hard probes 2012).

\bibitem{Agakishiev:2011dc}
G.~Agakishiev, et~al., {Identified hadron compositions in p+p and Au+Au
  collisions at high transverse momenta at $\sqrt{s_{_{NN}}} = 200$ GeV},
  Phys.Rev.Lett. 108 (2012) 072302.
\newblock \href {http://arxiv.org/abs/1110.0579} {\path{arXiv:1110.0579}},
  \href {http://dx.doi.org/10.1103/PhysRevLett.108.072302}
  {\path{doi:10.1103/PhysRevLett.108.072302}}.

\bibitem{Veldhoen}
M.~Veldhoen, these proceedings (hard probes 2012).

\bibitem{Davila}
A.~Daliva, these proceedings (hard probes 2012).

\bibitem{Tywunik}
K.~Tywoniuk, these proceedings (hard probes 2012).

\end{thebibliography}







\end{document}